\documentclass[%
 aip,
 amsmath,amssymb,
 reprint,%
]{revtex4-1}

\usepackage{graphicx}
\usepackage{dcolumn}
\usepackage{bm}
\usepackage[normalem]{ulem}
\usepackage[utf8]{inputenc}
\usepackage[T1]{fontenc}
\usepackage{mathptmx}
\usepackage{etoolbox}
\usepackage{braket}
\usepackage{xcolor}
\usepackage{comment}
\usepackage[colorlinks=true,
            linkcolor=blue,
            citecolor=blue,
            urlcolor=blue]{hyperref}
\makeatletter
\def\@email#1#2{%
 \endgroup
 \patchcmd{\titleblock@produce}
  {\frontmatter@RRAPformat}
  {\frontmatter@RRAPformat{\produce@RRAP{*#1\href{mailto:#2}{#2}}}\frontmatter@RRAPformat}
  {}{}
}%
\makeatother
\begin{document}
\preprint{AIP/123-QED}

\title{Chirality-Induced Spin Selectivity Regulates Triplet formation in Heliobacterial Photosynthesis}
\author{Parul Raghuvanshi}
\author{Vishvendra S. Poonia}
\email{vishvendra@ece.iitr.ac.in}
\affiliation{Department of Electronics and Communication Engineering, Indian Institute of Technology, Roorkee, India.}

\date{\today}

\begin{abstract}
Triplet formation and its regulation have been of central interest in understanding the photophysical behavior of living systems. In organic systems, excessive triplet formation poses significant challenges, as it can promote photochemical damage and reduce the efficiency of charge separation processes, making its regulation critically important. Here, we present a theoretical investigation of the intrinsic quantum spin dynamics governing triplet formation in the heliobacterial reaction center, a system that operates without any intrinsic magnetic interaction. Using an open quantum systems approach based on the Lindblad formalism, we simulate the spin-correlated radical pair dynamics occurring during charge separation in the heliobacterial reaction center. The study systematically examines the regulation of triplet formation through variations in recombination kinetics and hyperfine coupling strengths, and how this regulation is modified by the inclusion of chirality-induced spin selectivity (CISS) within the radical pair mechanism (RPM). Our results demonstrate that the CISS effect significantly suppresses triplet formation across the parameter space relevant to the heliobacterial molecular environment, revealing an intrinsic quantum protective mechanism operating through spin control in heliobacterial photosynthesis.

\end{abstract}

\maketitle
Photosynthesis, a natural biochemical process first termed in 1893 by Charles Reid Barnes and Conway MacMillan \cite{gest2002history}, sustains most life on Earth by producing oxygen and organic matter \cite{blankenship2021molecular}. Natural photosynthetic organisms such as plants, algae, and bacteria exhibit remarkably high solar energy conversion efficiency, with quantum yields of light-driven electron transfer in their reaction centers approaching unity \cite{blankenship2021molecular,hillier2001photosynthetic}. Therefore,understanding this highly efficient process has become a key focus of research aimed at addressing the growing global energy demand.Sunlight captured by antenna complexes funnels energy to the reaction center (RC), where charge separation generates electrochemical energy that drives downstream reactions into stable chemical fuel \cite{van2000photosynthetic}.\par
\begin{figure}[t]
\centering
\includegraphics[width=\linewidth]{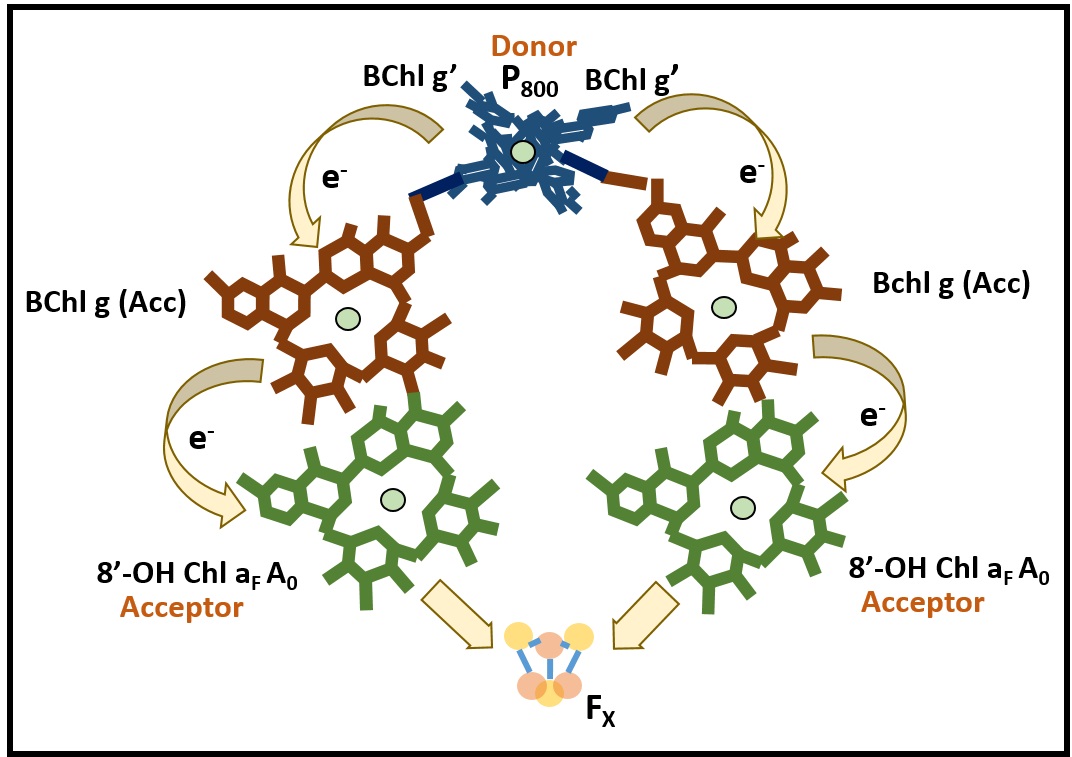}

\includegraphics[width=\linewidth]{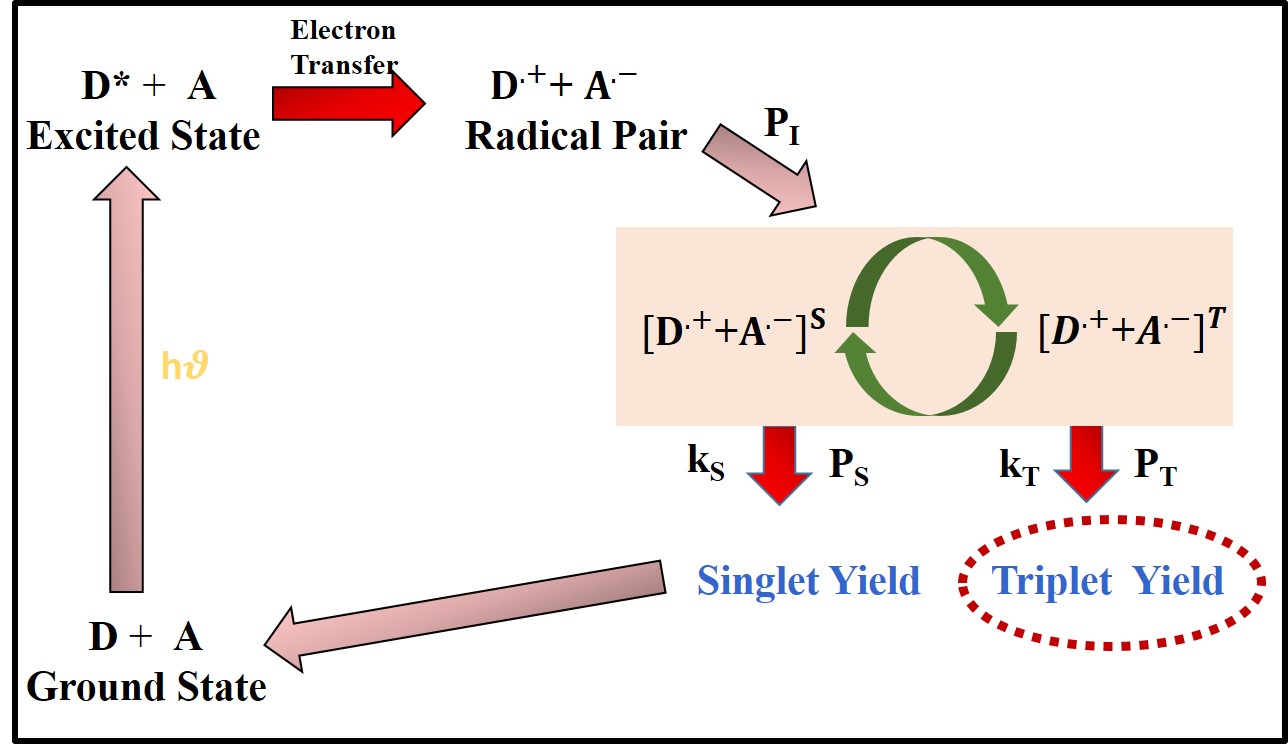}

\caption{Schematic of the heliobacterial reaction center and the associated radical-pair spin dynamics. The upper panel shows the sequential electron-transfer pathway from the donor ($D$) to the acceptor ($A$) through intermediate cofactors. The lower panel illustrates the CISS-assisted radical-pair model where $k_S$ and $k_T$ denote the singlet and triplet recombination rates, respectively. Red arrows indicate the role of CISS in regulating electron-transfer pathways and the resulting product yields.}
\label{fig:reaction_centre_rpm}
\vspace{-0.3cm}
\end{figure}

Heliobacteria is the simplest known anoxygenic photosynthetic bacteria identified  nearly a century later (1983) by Gest and Favinger \cite{gest1983heliobacterium} . Unlike other organisms with separate light-harvesting complexes, they bind antenna pigments directly to the core reaction center in a single unit \cite{amesz1995antenna,neerken2001antenna}.The structural simplicity of heliobacteria with fewer protein
subunits and a homodimeric design makes them particularly
attractive as model systems for understanding fundamental
photosynthetic mechanisms and as templates for developing
biomimetic artificial photosynthesis applications.The electron transfer cofactors for charge separation within the heliobacterial reaction center are  arranged into two equivalent branches\cite{gisriel2017structure,orf2018evolution}, each containing three specialized pigment molecules  as shown in Fig.~\ref{fig:reaction_centre_rpm}. Charge separation begins at the primary electron donor P800, a special pair formed by two bacteriochlorophyll g$'$ (BChl~g$^\prime$) molecules, with accessory BChl~g cofactors positioned intermediately to facilitate electron flow, finally reaching the primary electron acceptor A$_0$, which is 8$^1$-hydroxychlorophyll~a with a farnesyl side chain (8$^1$-OH–Chl~a$_\mathrm{F}$) \cite{gisriel2017structure,kim2024electronic}.During charge separation, radical-pair recombination or intersystem crossing can generate harmful chlorophyll triplet states\cite{monger1976triplet}. In most photosynthetic systems, these triplets are efficiently suppressed either by carotenoid-mediated quenching  \cite{alster2024direct}or by high-spin iron centers \cite{marais2015quantum} that reduce triplet yield. However, heliobacterial reaction centers lack high-spin iron and exhibit relatively slow triplet quenching combined with their high oxygen sensitivity \cite{kaur2023electronic,agostini2021differential}; this makes regulation of triplet formation particularly important in heliobacteria.\par

 The radical pair mechanism is a theoretical framework that has been widely employed to describe spin-dependent radical pair reactions in biological systems \cite{rodgers2009chemical,hore2016radical}. In our work, we use this framework to model the radical pair dynamics within the heliobacterial reaction center and to understand the regulation of triplet formation. This approach enables us to simulate both the coherent spin evolution and the relevant dissipative processes of the charge-separated state, thereby providing a direct route to calculate the quantum yield of the potentially harmful triplet state.Various studies shows that the well-established chiral-induced spin selectivity (CISS) effect, when considered along with the radical pair mechanism (RPM), significantly influences chemical outcomes in biological systems \cite{luo2021chiral,tiwari2024radical,tiwari2023quantum}.\par

In this work, we investigate the factors governing triplet yield formation in the heliobacterial reaction center through radical-pair dynamics. For this purpose,  we propose a radical-pair (RP) model in which bacteriochlorophyll g$^\prime$ acts as the donor and hydroxychlorophyll acts as the acceptor. In this model, initially the system is in the electronic ground state. Once photon energy is absorbed by the donor molecule, the donor becomes excited by promoting an electron to a higher-energy orbital and leaving behind a hole. The excited electron is then transferred via an accessory pigment molecule to the lowest unoccupied molecular orbital of the acceptor.Hence, a radical pair $(\mathrm{BChl}\,g^{\prime\bullet+} -  8^{1}\text{-OH--Chl}\,a_{F}^{\bullet-})$
is formed, in which one unpaired electron resides in a molecular orbital of the donor and the other unpaired electron resides in a molecular orbital of the acceptor. The correlated spin state of the radical pair evolves with time and undergoes inter-state transitions between the singlet and triplet states. The spin dynamics of the radical pair are governed by a Hamiltonian that captures the various interactions experienced by the system, as described in Eq.~\ref{eq:Hamiltonian}. After a certain time, the radical pair recombines and depending on its spin state at recombination, different reaction products are formed. The schematic of the RP model is shown in Fig.~\ref{fig:reaction_centre_rpm}.
\vspace{-0.2cm}
\begin{equation}
\begin{aligned}
\hat{H} ={}& \omega \left( \hat{S}_{Dz} + \hat{S}_{Az} \right)
+ \sum_{i=1}^{N_D} A_i \, \hat{\mathbf{I}}_i \cdot \hat{\mathbf{S}}_D
+ \sum_{j=1}^{N_A} A_j \, \hat{\mathbf{I}}_j \cdot \hat{\mathbf{S}}_A \\
& - J \, \hat{\mathbf{S}}_D \cdot \hat{\mathbf{S}}_A
+ \hat{\mathbf{S}}_D \cdot \mathbf{D} \cdot \hat{\mathbf{S}}_A .
\end{aligned}
\label{eq:Hamiltonian}
\end{equation}

Here, $\omega = g \mu_B B$ is the electron Larmor frequency, where $g$ is the
electron $g$-factor, $\mu_B$ is the Bohr magneton, and $B$ denotes the
external magnetic field. $\hat{\mathbf{S}}_D$ and $\hat{\mathbf{S}}_A$ are
the  spin operators of the donor and acceptor radicals
respectively, and A represents the hyperfine coupling tensor
associated with the $i$th and  $j$th nucleus. $J$ and $D$ denote the
interradical exchange interaction and the dipolar interaction 
respectively and $N_D + N_A = N$ is the total number of coupled nuclie.

\begin{figure*}[t]
\centering
\includegraphics[width=\textwidth]{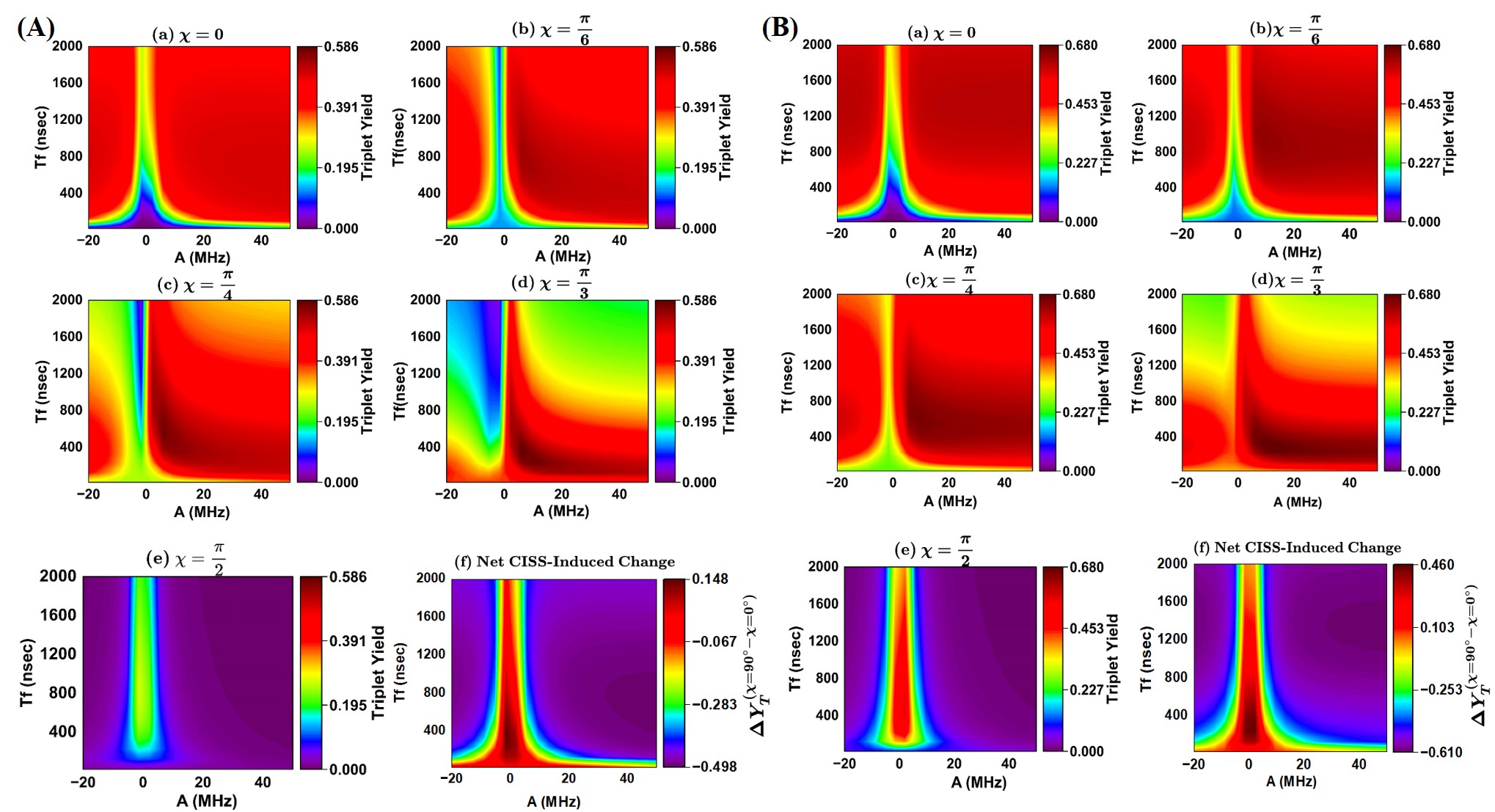}

\caption{Plots of the triplet yield as a function of the anisotropic hyperfine coupling $A$ (MHz) and the recombination time $T_f$ (ns) for the two-nuclei radical-pair model at different spin-selectivity angles, $\chi$. Panel (A) corresponds to the absence of an external magnetic field ($B=0$), while panel (B) corresponds to the presence of an external magnetic field of $B=50\,\mu\mathrm{T}$. For both cases, panels (a)--(e) present the triplet yield for $\chi=0^\circ$, $30^\circ$, $45^\circ$, $60^\circ$, and $90^\circ$, illustrating the influence of CISS effect on triplet-yield formation. Panel (f) in both cases shows the difference in the triplet yield, $\Delta Y_T=Y_T(\chi=90^\circ)-Y_T(\chi=0^\circ)$, highlighting the net CISS-induced change in the triplet yield.}

\label{fig:2N_aniso_combined}
\end{figure*}

The chirality-induced spin selectivity (CISS) effect influences both the
formation and recombination of the radical pair through spin-selective
electron transport within the chiral protein environment\cite{luo2021chiral}. Accordingly,
the impact of CISS is incorporated through the initial state $P_I$ and
the singlet and triplet recombination states $P_S$ and $P_T$ (see sec. 1.1 of \href{SupplementaryMaterial.pdf}{Supplementary Material}), which are
indicated by red arrows in Fig.~\ref{fig:reaction_centre_rpm}. 
The dynamics of the system are modeled using the quantum master equation approach described by Eq.~\ref{eq:master-equation}. Within this framework, we employ the concept of shelving states \cite{gauger2011sustained}, which are auxiliary states added to the Hilbert space to represent the final chemical products formed through spin-selective recombination of a radical pair. The spin-selective recombination is implemented using decay operators (see sec. 1.1 of \href{SupplementaryMaterial.pdf}{Supplementary Material}). These operators irreversibly transfer population from the electronic singlet and triplet subspaces into the corresponding shelving states $|S\rangle$ and $|T\rangle$. As recombination proceeds through one of these two channels, the final populations accumulated in the shelving states directly determine the singlet and triplet yields.
\begin{equation}
\begin{aligned}
\dot{\rho}
= {}& -\frac{i}{\hbar}\,[\hat{H},\rho]
+ k_S \left(
P_S \rho P_S^\dagger
- \frac{1}{2}\left(
P_S^\dagger P_S \rho
+ \rho P_S^\dagger P_S
\right)
\right) \\
& + k_T \left(
\sum_{T=t_0,t_\pm}
P_T \rho P_T^\dagger
- \frac{1}{2}\left(
P_T^\dagger P_T \rho
+ \rho P_T^\dagger P_T
\right)
\right).
\end{aligned}
\label{eq:master-equation}
\end{equation}
Here, $\rho$ is the state of the system and the solution of the master equation , $k_S$ and $k_T$ are the singlet and triplet recombination rates  respectively.\par

In this study, we focus on the key parameters of the radical pair model that directly influence the triplet yield and examine how different combinations of these parameters regulate triplet formation. We first examine the combined effects of hyperfine coupling and recombination time under the symmetric recombination condition ($k_S = k_T$) for five different values of the angle $\chi$ , where $\chi$ determines the degree of spin-selective electron transfer in the chiral medium . We further extend our analysis to asymmetric recombination rates ($k_S \neq k_T$), which represent a more biologically realistic scenario. The analysis is carried out for radical-pair systems containing  two (2N) nuclei in both the absence and presence of an external magnetic field. Our aim is to provide a comparative analysis of the triplet yield behavior for different values of $\chi$, examining how the yield varies with chirality across different regions of the parameter space.Furthermore, both isotropic and anisotropic cases are considered, but our primary focus is on the anisotropic case, as it represents a more physically realistic scenario capturing the full orientational dependence of the hyperfine interactions under actual physiological conditions. For completeness, the results for the isotropic case are included in the \href{SupplementaryMaterial.pdf}{Supplementary Material }.\par

Fig.~\ref{fig:2N_aniso_combined} shows the triplet yield as a function of the longitudinal component of the anisotropic hyperfine coupling constant, $A=A_{zz}$ (MHz), and the recombination time $T_f$ (ns), which is considered to be symmetric in this case, where $k_S=k_T=\frac{1}{T_f}$.The magnitude of the hyperfine coupling constant is scaled such that $\sqrt{N_1A_1}=\sqrt{N_2A_2}=A$, ensuring that the effective interaction strength varies with the number of nuclei coupled to the radical.  The results are presented for both in the absence of an external magnetic field [Fig.~\ref{fig:2N_aniso_combined}(A)] and in the presence of an external magnetic field of $50\mu\mathrm{T}$ [Fig.~\ref{fig:2N_aniso_combined}(B)]. For each case, calculations are performed for five spin-selectivity angles, $\chi=0,\ \pi/6,\ \pi/4,\ \pi/3,$ and $\pi/2$, corresponding to panels (a)--(e), respectively. The transverse components of the anisotropic hyperfine coupling are fixed at $A_{xx}=0.5417$ MHz and $A_{yy}=0.6769$ MHz~\cite{kim2024electronic}.The exchange and dipolar interaction strengths are fixed at $J=-0.20,\mathrm{mT}$ and $D=-0.17,\mathrm{mT}$, respectively.
 In Fig.~\ref{fig:2N_aniso_combined}(A), panel (a) corresponding to $\chi = 0^\circ$ (no CISS), the parameter space is predominantly occupied by high triplet-yield regions, as reflected by the extensive red areas spanning a wide range of hyperfine couplings and recombination times.In the narrow region around $A \approx -6$ to $+6$ MHz, the triplet yield decreases significantly and exhibits rapid variations with both hyperfine coupling and recombination time. A well-defined suppression zone emerges near zero hyperfine coupling, specifically within the range $A \approx -8$ to $+8$ MHz and for short recombination times below $T_f \approx 300$ ns, where the triplet yield approaches near-zero values. As the chirality-induced spin polarization increases (corresponding to $\chi=\pi/6$) in panel (b), the CISS effect begins to influence the spin dynamics. The maximum triplet yield rises slightly at moderate recombination times  and small positive hyperfine couplings . More notably, the triplet yield within the narrow hyperfine-coupling range ($-6$ to $+6$ MHz) decreases,as indicated by the sky-blue region. As the spin polarization is further increased (corresponding to $\chi=\pi/4$, panel(c)), the maximum triplet yield slightly increase  , occurring for hyperfine couplings in the range $A\approx2$--13 MHz and recombination times around $T_f\approx500$--860  ns. Upon further increasing the spin polarization (corresponding to $\chi=\pi/3$, panel (d)), the influence of spin polarization due to chirality on the spin dynamics becomes most pronounced. This is evident from the substantial change in the triplet-yield landscape compared to the lower spin-selectivity angles.. Finally, in panel (e), corresponding to $\chi=\pi/2$ (full CISS ), the triplet yield is strongly suppressed across most of the parameter space. The predominance of deep-purple regions indicates that the triplet yield approaches near-zero values over a wide range of hyperfine couplings and recombination times.
\newline To examine how an external magnetic field affects the regulation of triplet formation in helioabcteria, we introduce a field with strength $B_0$ whose direction can vary relative to the radical pair (RP). We express this magnetic field in spherical coordinates as $\mathbf{B} = B_0 (\sin \theta \cos \phi \, \hat{x} + \sin \theta \sin \phi \, \hat{y} + \cos \theta \, \hat{z})$. The angles $\theta$ and $\phi$ specify the orientation of the magnetic field with respect to the hyperfine axes \cite{gauger2009quantum} (see sec.2 of \href{SupplementaryMaterial.pdf}{ Supplementary Material}) . To model realistic scenarios in which the RP can assume arbitrary orientations relative to the applied field, we calculate the angular-averaged triplet yield $Y_{T,\mathrm{Avg.}}$ using Eq.~\eqref{eq:angular_average} for each value of $B_0$. The averaging is performed over all possible orientations, with $\theta \in [0,\pi]$ and $\phi \in [0,2\pi]$.
\begin{equation}
Y_{T,\mathrm{Avg.}} = \frac{1}{4\pi}
\int_{0}^{2\pi}
\int_{0}^{\pi}
Y_T \sin\theta \, d\theta \, d\phi
\label{eq:angular_average}
\end{equation}

Fig.~\ref{fig:2N_aniso_combined}(B) shows the triplet-yield behavior in the presence of an external magnetic field of $50\,\mu\mathrm{T}$, with all other parameters kept the same as in the zero-field case. In this case, the plotted triplet yield corresponds to the average triplet yield calculated using Eq.~(\ref{eq:angular_average}). The results indicate that the application of the external magnetic field enhances the overall triplet yield compared to the field-free case. Despite this increase, the overall triplet-yield landscape remains qualitatively preserved. In particular, for $\chi=0$ to $\chi=\pi/3$ [panels (a)--(d)], the contour patterns closely resemble those observed in the absence of the magnetic field, indicating that the external field primarily increases the triplet-yield magnitude (reaching 0.680 from 0.586) without significantly altering the underlying dependence on the hyperfine coupling and recombination time. For $\chi=\pi/2$ [panel (e)] (full CISS case), the triplet yield is strongly suppressed across most of the parameter space. Since  the triplet-yield varies widely across the parameter space, the overall influence of CISS is not clearly evident. To better visualize this effect, we plot the difference in triplet yield between the no-CISS case ($\chi=0$) and the full-CISS case ($\chi=\pi/2$) in Fig.~\ref{fig:2N_aniso_combined}(f). A maximum reduction of approximately 49.8\% in the triplet yield is observed as the system evolves from the no-CISS to the full-CISS case in the absence of an external magnetic field [Fig.~\ref{fig:2N_aniso_combined}(A-f)]. In the presence of an external magnetic field [Fig.~\ref{fig:2N_aniso_combined}(B-f)],  the reduction induced by CISS becomes even more pronounced, reaching approximately 61\%. In addition, the differences between consecutive spin-selectivity angles are provided in  sec.3 of \href{SupplementaryMaterial.pdf}{Supplementary Material}. These difference plots not only confirm that CISS regulates the triplet yield but also reveal the nature of this regulation. Specifically, the predominantly negative values of $\Delta Y_T$ indicate that the primary role of chirality within the radical-pair mechanism is to suppress triplet formation in heliobacterial photosynthesis.\par
\begin{figure}[htbp]
\centering
\includegraphics[width=\columnwidth]{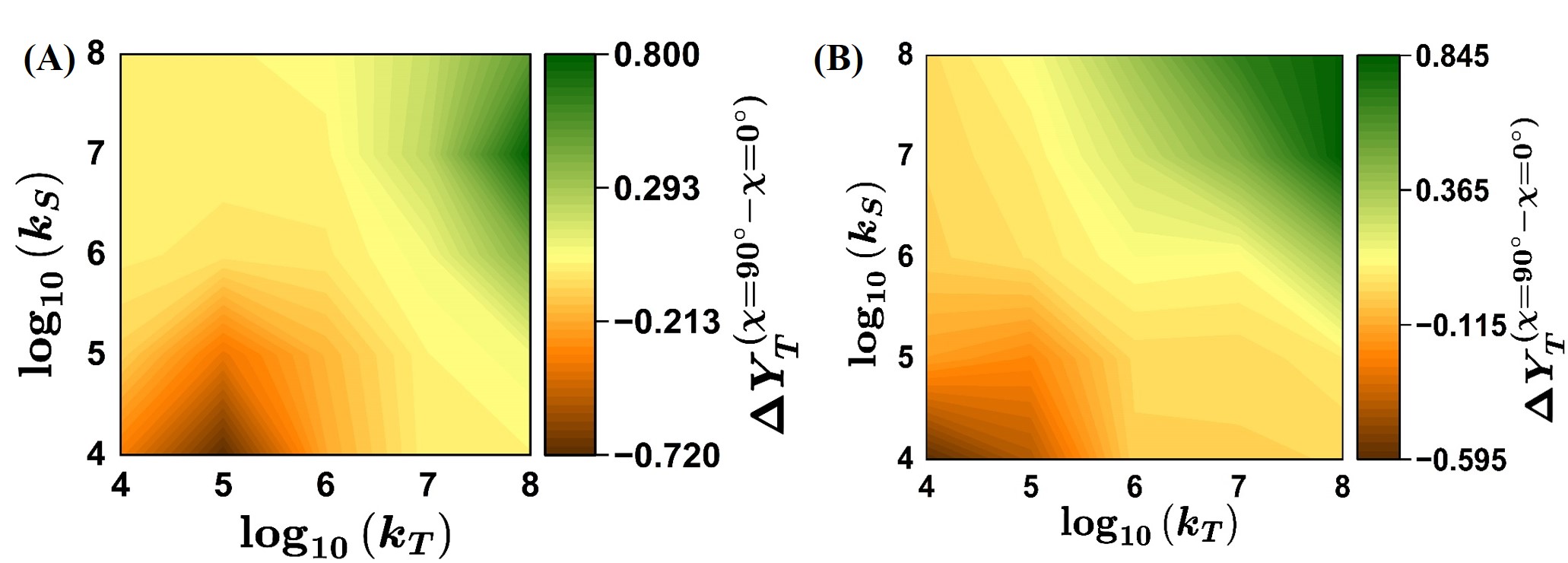}
\caption{Plots of the difference in the triplet yield, $\Delta Y_T = Y_T(\chi=90^\circ)-Y_T(\chi=0^\circ)$, illustrating the net influence of CISS on triplet yield formation. The quantity $\Delta Y_T$ is plotted as a function of the singlet recombination rate $k_S$ and triplet recombination rate $k_T$. Panel (A) corresponds to $B=0$, while panel (B) corresponds to $B=50\,\mu\mathrm{T}$.}
\label{fig:rATE_Difference Plot}
\end{figure}
To further investigate the influence of chirality-induced spin polarization, we performed a rate analysis by plotting the triplet yield as a function of the singlet and triplet recombination rates, $\log_{10}(k_S)$ and $\log_{10}(k_T)$, respectively, for different spin-selectivity angles, $\chi$. Particular emphasis is placed on the biologically relevant asymmetric recombination regime ($k_S>k_T$) \cite{kattnig2016electron,tiwari2024radical,denton2024magnetosensitivity}, which is consistent with recombination rates reported for heliobacterial reaction centers \cite{kim2024electronic}. The complete parameter set and corresponding results for different values of $\chi$ are provided in sec.3.1 of  \href{SupplementaryMaterial.pdf}{Supplementary Material}. The overall effect of spin polarization due to chirality is summarized in Fig.~\ref{fig:rATE_Difference Plot}.
In the absence of a magnetic field [Fig.~\ref{fig:rATE_Difference Plot}(A)], the largest suppression (21--72\%) occurs in the symmetric recombination regime ($10^{4}\leq k_S, k_T\leq10^{6}$). In the biologically relevant regime ($k_S>k_T$), the suppression is about 3\% for $k_S=10^{6}$ and $k_T=10^{5}$, which further decreases to approximately 0.3--0.7\% as $k_S$ increases.In the presence of a magnetic field [Fig.~\ref{fig:rATE_Difference Plot}(B)], the suppression is weaker, with a maximum reduction of about 59.5\% in the symmetric regime. In the $k_S>k_T$ region, only a very small suppression (0.01--0.05\%) is observed. Overall, these results show that spin polarization induced by chirality suppresses triplet formation even in the biologically relevant asymmetric recombination regime..\par
In summary, we investigate the role of chirality-induced spin selectivity in regulating triplet yield formation in the heliobacterial reaction center using a radical-pair model. We observe that the triplet yield is highly sensitive to the  angle $\chi$, which parametrizes the degree of spin-selective electron transfer through a chiral medium. As $\chi$ increases from $0$ to $\pi/2$, a consistent and pronounced suppression of the triplet yield is observed across the explored parameter space. This suppression behavior remains qualitatively robust even in the presence of an external magnetic field. Although the magnetic field raises the absolute magnitude of the triplet yield, but the CISS-induced suppression is nevertheless preserved,demonstrating the robustness of the observed trends with and without an external magnetic field..We further observe that the weak hyperfine coupling region ($-6\,\mathrm{MHz}$ to $6\,\mathrm{MHz}$) emerges as a critical window where the triplet yield varies rapidly, indicating high sensitivity of the spin dynamics. This behavior arises from the near-degeneracy of the singlet and triplet states, which enhances singlet--triplet mixing and makes the system highly responsive to small perturbations. Consequently, the radical pair rapidly oscillates between the singlet and triplet states rather than remaining in a single spin state. Overall, our results suggest that the CISS effect plays a quantum protective role in heliobacterial photosynthesis by systematically suppressing triplet formation across the biologically relevant parameter space, thereby potentially safeguarding the photosynthetic apparatus against the generation of harmful reactive oxygen species.In the present study, we have restricted our analysis to systems with up to two nuclear spins, which is sufficient to capture the essential physics of the radical pair mechanism in heliobacteria. Nevertheless, the present work can be extended in several directions. Future studies may consider systems with a larger number of nuclear spins to investigate the spin dynamics more comprehensively and to assess the scalability of the CISS-induced triplet suppression. Additionally, the incorporation of further spin interaction parameters, such as exchange and dipolar coupling, would allow a more complete exploration of their influence on spin dynamics and triplet yield suppression. It would also be of considerable interest to examine the role of environmental noise and quantum decoherence in conjunction with the CISS effect, in order to investigate how chirality-induced spin selectivity contributes to the protective mechanism in heliobacteria within a larger and more realistic molecular environment.

\section*{Acknowledgments}
The authors gratefully acknowledge financial support from the National Quantum Mission (NQM) of the Department of Science and Technology (DST), Government of India, and the Ministry of Electronics and Information Technology (MeitY) through Grant Nos. DST/QTC/NQM/QC/2024/1 and 4(3)/2024-ITEA. We sincerely thank Ashutosh Tripathi for his valuable support throughout this work. We dedicate this work to the memory of our late research group member, Jothishwaran C.A., who passed away during the course of this study. His brilliance, encouragement, and unwavering support to our research group were profoundly influential and will always be remembered.

\makeatletter
\def\bibsection{\section*{REFERENCES}}
\makeatother
\nocite{*}
\bibliographystyle{aipnum4-1} 
\bibliography{aipsamp}

\end{document}